\documentclass[11pt]{article}
\usepackage[top=1in,bottom=1in,left=1in,right=1in]{geometry}
\pdfoutput = 1
\usepackage[noEucal]{main}
\usepackage{aas_macros}
\usepackage[utf8]{inputenc}
\makeatletter \g@addto@macro\@floatboxreset\centering \makeatother
\usepackage{graphicx}
\usepackage[scr=boondoxo]{mathalfa}
\usepackage[bbgreekl]{mathbbol}
\DeclareSymbolFontAlphabet{\mathbbm}{bbold}
\usepackage[font={small,it}]{caption}

\def\soft{{soft}}

\def\soft{\text{soft}}
\def\eff{{\text{eff}}}
\def\inter{{\text{int}}}

\begin{document}
\begin{titlepage}
\unitlength = 1mm

\hfill CALT-TH 2024-012

\vskip 3cm
\begin{center}

\openup .5em

{\huge{An On-Shell Derivation of the \\ Soft Effective Action in Abelian Gauge Theories}}

\vspace{0.8cm}
Temple He$^\ddagger$, Prahar Mitra$^{\Diamond}$, Allic Sivaramakrishnan$^\ddagger$, Kathryn M. Zurek$^\ddagger$

\vspace{1cm}

{\it  $^\ddagger$Walter Burke Institute for Theoretical Physics, California Institute of Technology, \\ Pasadena, CA 91125 USA}\\
{\it  $^\Diamond$Institute for Theoretical Physics, University of Amsterdam,
Science Park 904, Postbus 94485, 1090 GL Amsterdam, The Netherlands}

\vspace{0.8cm}

\begin{abstract}

We derive the soft effective action in $(d+2)$-dimensional abelian gauge theories from the on-shell action obeying Neumann boundary conditions at timelike and null infinity and Dirichlet boundary conditions at spatial infinity. This allows us to identify the on-shell degrees of freedom on the boundary with the soft modes living on the celestial sphere. Following the work of Donnelly and Wall, this suggests that we can interpret soft modes as entanglement edge modes on the celestial sphere and study entanglement properties of soft modes in abelian gauge theories.

\end{abstract}

\vspace{1.0cm}
\end{center}
\end{titlepage}
\pagestyle{empty}
\pagestyle{plain}
\pagenumbering{arabic}

\tableofcontents

\section{Introduction}

The infrared (IR) sector of quantum field theories (QFTs) has recently enjoyed much attention, primarily due to the seminal work of Strominger \cite{Strominger:2013jfa, He:2014laa}, which showed that Weinberg's leading soft graviton theorem \cite{Weinberg:1965aa} is the Ward identity for the BMS supertranslation symmetry \cite{Bondi:1962px, Sachs:1962wk}. This relationship between soft theorems and the so-called \emph{asymptotic symmetries} is now understood to be a universal feature of all gauge and gravitational theories in asymptotically flat spacetimes (see \cite{Strominger:2017zoo, Pasterski:2021rjz, Raclariu:2021zjz} for a review). For example, the leading soft photon or soft gluon theorems in gauge theories \cite{Weinberg:1965aa, BERENDS1989595} are the Ward identity for large gauge symmetries \cite{Strominger:2013lka, He:2014cra, He:2015zea, Campiglia:2015qka, Kapec:2014zla, Kapec:2015ena, Nande:2017dba, He:2019jjk, He:2020ifr, He:2023bvv, Freidel:2023gue}, and the subleading soft graviton theorem in gravitational theories \cite{Cachazo:2014fwa} is the Ward identity for BMS superrotations \cite{Barnich:2011ct, Kapec:2014opa}.

A particularly important consequence of these symmetries is that gauge and gravitational theories do not have a unique vacuum state. Rather, they have infinitely many vacua, all related via action by the asymptotic symmetry charge. In other words, the asymptotic symmetry is spontaneously broken by the choice of vacuum state.\footnote{The vacuum degeneracy being discussed here is \emph{not} the one associated with the $\t$-angle in gauge theories, which is related to gauge transformations that are constant on the boundary but are topologically non-trivial (i.e., they have a non-zero winding number). We are interested in degeneracy due to gauge transformations that are non-constant on the boundary but are topologically trivial.} The Goldstone mode $\t(x)$ associated with this spontaneous breaking lives on the celestial sphere $\mss^d$, the codimension-two boundary of the spacetime (the bulk spacetime dimension is $d+2$). The symplectic conjugate of the Goldstone mode is the so-called \emph{soft photon operator} $\phi(x)$, which inserts a soft (low energy) photon in a scattering amplitude \cite{He:2020ifr}. Together, the fields $\t,\phi$ constitute the low energy sector of the theory. They live on a codimension-two boundary of the spacetime and, in this sense, are the boundary or edge modes of abelian gauge theories.

The effective dynamics of the edge modes are described by a codimension-two action which was constructed in \cite{Kapec:2021eug} (similar actions in various other forms have previously appeared in \cite{Himwich:2020rro, Arkani-Hamed:2020gyp, Magnea:2021fvy, Gonzalez:2021dxw,  Kalyanapuram:2021tnl} as well, though they all exclusively work in four spacetime dimensions). The so-called \emph{soft effective action} reproduces all universal soft features of abelian gauge theories, namely Weinberg's leading soft photon theorem \cite{Weinberg:1965aa} and the IR factorization of scattering amplitudes (IR divergences in four dimensions) \cite{YENNIE1961379}. More precisely, given a gauge theory with IR cutoff $\mu$, and denoting the energy scale separating the soft modes from the hard ones as $\L$, scattering amplitudes take the form
\begin{align}
\label{intro:soft_factor}
\avg{ \phi(x_1) \cdots \phi(x_m) \CO_1 \cdots \CO_n }_\mu = \left( \CJ(x_1) \cdots \CJ(x_m)  e^{-\G(\mu,\L)}  \right) \avg{ \CO_1 \cdots \CO_n }_\L ,
\end{align}
where $\la \cdots \ra_E$ denotes a scattering amplitude evaluated with IR cutoff $E$, $\CO_k$ are the hard insertions (energy above the scale $\L$), $\phi(x_i)$ are the soft photon insertion (with energy below $\L$ but above $\mu$), $\CJ(x_i)$ are related to the leading Weinberg soft factor, and $\G(\mu,\L)$ is the contribution from virtual soft photons. Thus, we see the amplitude factorizes into a ``hard amplitude'' and a soft factor that receives contributions from virtual soft photons and external soft photons. When $\mu$ and $\L$ are small compared to all the other energy scales in the amplitude, the soft factor is universal and can be reproduced by a path integral of the form
\begin{align}
\label{intro:soft_pi}
\int [ \dt \phi ] [\dt \t ] \, e^{-S_\eff[\phi,\t]} \phi(x_1) \cdots \phi(x_m) = \CJ(x_1) \cdots \CJ(x_m)e^{-\G[\mu,\L]} ,
\end{align}
where the effective action $S_\eff[\phi,\t]$ for the soft or edge modes is given by (see Section \ref{sec:sea} for details)
\begin{equation}
\begin{split}
\label{intro:sea}
    S_{\eff} [\phi,\t] =  \a \int_{\mss^d} \frac{ \dt^{d} x}{(2\pi)^{d}} \big( \p_a \phi (x) \big)^2 - \frac{i}{2c_{1,1}} \int _{\mss^d} \dt^{d} x \, \wt{\p^a\t}(x) \big( \p_a \phi (x) - \p_a \CJ(x) \big)  ,
\end{split}
\end{equation}
where $\mss^d$ is the celestial sphere. Interestingly, this action is neither real nor local (the tilde superscript denotes the shadow transform \eqref{shadow_def}, which is a non-local integral transform). It was constructed in \cite{Kapec:2021eug} using the asymptotic symmetries of the theory (in this case, large gauge transformations) and relied heavily on effective field theory techniques. Therefore, it is interesting to ask whether the action \eqref{intro:sea} involving soft modes can be derived directly from the bulk action, whose on-shell degrees of freedom are the edge modes.

Edge modes in gauge theories were introduced to the study of entanglement entropy by Donnelly \cite{Donnelly:2011hn}. It is known from \cite{Kabat:1995eq} that the entanglement entropy of Maxwell theory in $d+2$ dimensions is equal to that of $d$ scalar fields plus an additional contact term, whose physical significance was not clear. In \cite{Donnelly:2014fua, Donnelly:2015hxa}, Donnelly and Wall showed that this contact term \emph{is} physical and is, in fact, the contribution of edge modes to the entanglement entropy. They further showed that the effective action for the edge modes could be obtained by evaluating the Maxwell action on-shell with ``magnetic conductor boundary conditions.'' These ``entanglement edge modes'' also live on a codimension-two boundary of the spacetime (specifically, on the entangling surface), and given their remarkable similarity to the ``soft edge modes'' on the celestial sphere appearing in \eqref{intro:sea}, it is reasonable to expect that they are in fact related (for example, see \cite{Chen:2023tvj}). In this paper, we prove that this is indeed the case by showing that the soft contribution to the on-shell action of abelian gauge theories is \emph{exactly} equal to the soft effective action \eqref{intro:sea}. There are, however, two crucial ways in which our setup differs from that of Donnelly and Wall in \cite{Donnelly:2014fua,Donnelly:2015hxa}.

Firstly, as was remarked previously, Donnelly and Wall imposed magnetic conductor boundary conditions, which fixes $B_\|=0$ and $E_\perp$, so that the effective action for the entanglement edge modes is a function of $E_\perp$.\footnote{The entanglement entropy of the entanglement edge modes is then determined by evaluating the path integral over the modes $E_\perp$.} On the other hand, the soft degrees of freedom that we are interested in live on a cut of asymptotic null infinity $\CI^\pm$, so we will instead impose Neumann boundary conditions, which allows non-trivial radiation flux through the boundary. This requires us to add extra boundary terms similar to Gibbons-Hawking-York (GHY) terms in general relativity to the Maxwell action, so that
\begin{equation}
\begin{split}
\label{intro:S_action}
S[A,\Phi] = S_{\CM}[A,\Phi] + \frac{1}{e^2} \int_{\S^+} A \w \star F - \frac{1}{e^2} \int_{\S^-} A \w \star F ,
\end{split}
\end{equation}
where $S_\CM$ is the bulk action (including matter fields $\Phi$), and $\Sigma^\pm \equiv \CI^\pm \cup i^\pm$ are the non-spacelike boundaries. Secondly, the edge mode contribution to the entanglement entropy studied in \cite{Donnelly:2014fua,Donnelly:2015hxa} is an ultraviolet (UV) effect, which arises from degrees of freedom living close to the entangling surface and is dealt with in the usual way through renormalization. However, in our analysis, since the surface of interest lives on the \emph{asymptotic} boundary of spacetime, we have to deal with additional IR divergences (at least in four dimensions). Therefore, we must be more careful about how to evaluate the bulk part of the action \eqref{intro:S_action} on-shell, and appropriately determine the $i\e$ prescription in the Lorentzian path integral.\footnote{
This was not an issue for Donnelly and Wall in \cite{Donnelly:2014fua,Donnelly:2015hxa}, as their entanglement entropy calculation was done in Euclidean signature.} Our goal is to show that once these subtleties are dealt with, the relation between the soft contribution to the on-shell action and the soft effective action is given by
\begin{align}\label{final-result0}
\begin{split}
S[A,\Phi]\big|_{\soft\text{+on-shell}} = i S_\eff[\phi,\theta] ,
\end{split}
\end{align}
where the extra factor of $i$ is present due to the fact that $S[A,\Phi]$ is a Lorentzian action whereas $S_\eff$ is Euclidean \eqref{intro:soft_pi}. This is the main result of our work. 

Because the entanglement edge modes that are studied by Donnelly and Wall \cite{Donnelly:2014fua, Donnelly:2015hxa}, albeit using different boundary conditions, are precisely the on-shell modes living on a codimension-two boundary, our result \eqref{final-result0} solidifies the connection between the entanglement edge modes and the soft modes obtained from a symplectic analysis \cite{He:2020ifr, He:2023bvv}. This is perhaps not entirely surprising, as it is natural in many regards to identify the soft modes with entanglement edge modes, both of which live on codimension-two surfaces. Nevertheless, we view the novel feature in our analysis to be the determination of precisely which boundary conditions allow us to establish an equivalence between the two types of boundary modes.

Relating soft and entanglement edge modes in gauge theory lays the foundation for doing the same in gravity. Gravitational edge modes enter into the study of subregions in gravity, where they help answer the question: What are the degrees of freedom associated with a subregion in gravity? We therefore anticipate that applying our approach to gravity may connect soft modes in gravity to entanglement edge modes and, in turn, to objects utilized to diagnose entanglement, such as the modular Hamiltonian proposed in \cite{Verlinde:2022hhs}.  For instance, by determining the appropriate GHY boundary terms needed such that the soft limit of the on-shell action reproduces the soft effective action in gravity, we may conclude that the corresponding boundary conditions for the gravitational edge modes are a ``natural'' choice. We leave such directions for future work.

This paper is organized as follows. We will introduce the preliminaries involving soft theorems and soft factorization of amplitudes in Section~\ref{sec:prelim}. In Section~\ref{sec:edge-mode}, we will perform the computation that establishes the equivalence between the on-shell action capturing the edge mode degrees of freedom and the soft effective action. We summarize our results in Section~\ref{sec:summary}. In Appendix~\ref{app:current-shadow}, we prove a technical identity that relates the matter current to its shadow transform, which is instrumental in showing the equivalence between the two actions. In Appendix~\ref{app:massive}, we also take into account massive matter particles that may be present in the theory.

\section{Preliminaries}\label{sec:prelim}

We begin by establishing the necessary prerequisites. In Section~\ref{ssec:notation}, we introduce the notation and conventions used throughout this paper. We will be following those given in Appendix A of both \cite{He:2019jjk} and \cite{He:2023bvv}, where more details can be found. In Section~\ref{ssec:soft-factor}, we present a brief review of soft factorization in scattering amplitudes and introduce the soft effective action derived in \cite{Kapec:2021eug}.

\subsection{Notations and Conventions}\label{ssec:notation}

\paragraph{Position Space Coordinates:} 

Our theory lives in $(d+2)$-dimensional Minkowski spacetime, $\CM = \mrr^{1,d+1}$, and for computational simplicity we will work in flat null coordinates $x^\mu = (u,x^a,r)$, where $u,r\in\mrr$ and $x^a \in \mrr^d$. These are related to Cartesian coordinates $X^A$ by
\begin{equation}
\begin{split}
\label{flat_null}
X^A = r {\hat q}^A(x) + u n^A , \qquad {\hat q}^A(x) = \left( \frac{1 + x^2}{2} , x^a , \frac{1 - x^2}{2} \right) , \qquad n^A = \left( \frac{1}{2} , 0^a , - \frac{1}{2} \right) .
\end{split}
\end{equation}
Note that ${\hat q}^A(x)$ and $n^A$ are null and $n \cdot {\hat q}(x) = - \frac{1}{2}$. It follows the Minkowski line element in flat null coordinates is given by
\begin{equation}
\begin{split}
\label{Minkowski_metric}
\dt s^2 = \eta_{AB} \, \dt X^A \,\dt X^B = -\dt u \, \dt r + r^2 \d_{ab}\, \dt x^a \, \dt x^b .
\end{split}
\end{equation}
We will throughout this paper use lowercase Greek letters $\mu,\nu,\ldots$ to denote flat null coordinates and capital Latin letters $A, B,\ldots$ to denote Cartesian coordinates. Lowercase Latin indices denote the transverse directions along the celestial sphere $\mss^d$ and are raised and lowered by the Cartesian metric $\delta_{ab}$. 

The null boundaries $\CI^\pm$ are located at $r \to \pm \infty$ while keeping $(u,x)$ fixed, and their topology is given by $\mrr \times \mss^d$. The past (future) boundary of $\CI^+$ ($\CI^-$) is located at $u =-\infty$ ($u=+\infty$) and is denoted by $\CI^+_-$ ($\CI^-_+$). The point labeled by coordinate $x^a$ on $\CI^+$ is antipodal to the point with the same coordinate value on $\CI^-$.\footnote{The antipodal point on $\mss^d$ can be defined by embedding $\mss^d \hookrightarrow \mrr^{d+1}$, which maps $x^a \mapsto \vec{X}(x)$, where $\vec{X} \cdot \vec{X} = 1$. On $\mrr^{d+1}$, the antipodal map is given by $\vec{X} \mapsto - \vec{X}$.} In these coordinates, the integration of forms on $\CM$, $\CI^\pm$ and $\CI^\pm_\mp$ are given by (we follow the conventions outlined in Appendix A of \cite{He:2023bvv})
\begin{equation}
\begin{split}
\label{volume_forms}
    \int_\CM C_{d+2} &= - \frac{1}{2} \int_{\mrr} \dt u \int_\mrr \dt r \int_{\mss^d}\dt^d x \, |r|^d ( \star C_{d+2} )  , \\
\int_{\CI^\pm} C_{d+1} &= - \frac{1}{2} \int_\mrr \dt u \int_{\mss^d} \dt^d x \left( \lim_{r \to \pm \infty} |r|^d ( \star C_{d+1} )^r \right) , \\
\int_{\CI^\pm_\mp} C_d &= \frac{1}{2} \int_{\mss^d} \dt^d x  \left( \lim_{u \to \mp \infty}  \lim_{r \to \pm \infty}  |r|^d ( \star C_d)^{ur} \right) ,
\end{split}
\end{equation}
where $C_p$ denotes a $p$-form.

\paragraph{Momentum Space Coordinates:} An off-shell momentum is parametrized by
\begin{equation}
\begin{split}
\label{offshell_mom}
    \ell^A = \o \big( {\hat q}^A(x) + \k n^A \big) ,
\end{split}
\end{equation}
where $\ell^2= - \kappa \omega^2$. The off-shell integration measure is 
\begin{equation}
\begin{split}
\label{offshell_int}
\int_\CM \frac{\dt^{d+2} \ell}{(2\pi)^{d+2}} = \frac{1}{4\pi} \int_{|\o|>\mu} \dt \o \, |\o|^{d+1} \int_{\mss^d} \frac{\dt^d x}{(2\pi)^d}   \int_\mrr \frac{\dt \k}{2\pi} ,
\end{split}
\end{equation}
where to deal with IR divergences, all momentum space integrals are performed with a cut-off $\mu$, which is taken to be much smaller than all other scales in the problem. Similarly, we use the following parametrization for on-shell momenta:
\begin{equation}
\begin{split}
\label{mompar}
    p^A  = \o \left[ {\hat q}^A(x) + \left(\frac{m^2}{\o^2}\right) n^A \right] , \qquad p^2 = - m^2 .
\end{split}
\end{equation}
The properties and advantages of using the flat null coordinates for position and momenta were further expounded in Appendix A of \cite{He:2019jjk}.

\paragraph{Scattering Amplitudes:} 

Given an IR cutoff $\mu$, an $n$-point scattering amplitude can be written as a time-ordered vacuum correlation function, such that
\begin{equation}
\begin{split}
\label{amp_def}
A_n = \avg{ \CO_1 \cdots \CO_n }_\mu ,
\end{split}
\end{equation}
where we denoted\footnote{This definition for $\CO_k$ is the one implemented by the LSZ reduction formula (see Section 4.3 of \cite{He:2020ifr}).}
\begin{equation}
\begin{split}
\label{Ok_def}
\CO_k \equiv \t ( \o_k ) \Big[ \CO^+_k (\o_k {\hat q}(x_k)) - \CO^-_k (\o_k {\hat q}(x_k))  \Big]  + \t(-\o_k) \Big[ \CO^{-}_k(-\o_k{\hat q}(x_k))^\dag - \CO^{+}_k(-\o_k{\hat q}(x_k))^\dag  \Big]  . 
\end{split}
\end{equation}
Here, $\t(\o)$ is the Heaviside step function, the $\pm$ superscript corresponds to either the outgoing ($+$) or incoming ($-$) mode, and $\CO_k^{\pm}$ ($\CO_k^{\pm\dagger}$) is the annihilation (creation) operator for the $k$th particle. Furthermore, we denote the operator that inserts a photon with momentum $q^A = \o {\hat q}^A(x)$ and polarization $a$ by $\CO_a(\o,x)$. The corresponding polarization vector is given by
\begin{equation}
\begin{split}
\label{polpar}
\ve^A_a(x) = \p_a {\hat q}^A(x) = \big( x_a , \d_a^b , - x_a \big) . 
\end{split}
\end{equation}

\subsection{Soft Factorization}
\label{ssec:soft-factor}

\subsubsection{Real Soft Photons}

Weinberg's leading soft photon theorem \cite{Weinberg:1965aa} states that a scattering amplitude with $m$ photons, each with momentum $q_i$ and polarization $a_i$, and $n$ hard particles, each with momentum $p_k$ and $U(1)$ charge $Q_k \in \mzz$, factorizes in the leading soft limit ($q_i^0 \ll p_k^0$ for all $i$ and $k$) as\footnote{In abelian gauge theories, the soft limit can be taken either consecutively or simultaneously without any ambiguity. This is no longer the case for nonabelian gauge theories.}
\begin{equation}
\begin{split}
\label{soft_thm_1}
A_{m+n} \quad &\xrightarrow{q_i \to 0} \quad  \SS_m^\0 A_n , \qquad \SS_m^\0 \equiv \prod_{i=1}^m \left( e \sum_{k=1}^n Q_k \frac{p_k \cdot \ve_{a_i}(q_i) }{ p_k \cdot q_i - i \e} \right) ,
\end{split}
\end{equation}
where the superscript on $\SS_m^\0$ signifies this is the \emph{leading} soft factor. To recast this into a cleaner form, we will utilize the notation introduced in the previous subsection. First, we define the soft photon operator 
\begin{equation}
\begin{split}\label{Na_def}
N_a(x) \equiv \frac{1}{2e} \left( \lim_{\o \to 0^+} + \lim_{\o \to 0^-} \right) \big( \o \CO_a(\o,x) \big)  = N_a^+(x) - N_a^-(x) ,
\end{split}
\end{equation}
where $N_a^\pm(x)$ are the Hermitian $out$ and $in$ soft photon operators \cite{He:2020ifr},\footnote{In \eqref{Napm_def}, we have used $\t(0)=\frac{1}{2}$. This choice is not arbitrary and is rather precisely what we get if we use the LSZ representation of the operators defined in \eqref{Ok_def} and then take the limit in \eqref{Na_def} (see Section of 4.3 of \cite{He:2020ifr}). }
\begin{equation}
\begin{split}
\label{Napm_def}
    N_a^\pm(x) \equiv \lim_{\o \to 0^+} \frac{1}{e}  \o \CO^{\pm}_a (\o {\hat q}(x))   = N_a^{\pm}(x)^\dag  . 
\end{split}
\end{equation}
Note that the factor of $\o$ is needed to cancel the simple pole in the soft factor at $q_i=0$ so that the soft limit is well-defined. Using this, \eqref{soft_thm_1} can be written as \cite{Kapec:2021eug}
\begin{equation}
\begin{split}\label{soft_thm}
\avg{ N_{a_1}(x_1) \cdots N_{a_m}(x_m) \CO_1 \cdots \CO_n }_\mu = \CJ_{a_1}(x_1) \cdots \CJ_{a_m}(x_m) \, \avg{ \CO_1 \cdots \CO_n }_\mu ,
\end{split}
\end{equation}
where
\begin{equation}
\begin{split}
\label{Ja_def}
\CJ_a(x) \equiv \p_a \sum_{k=1}^n Q_k \ln | p_k \cdot {\hat q}(x) | = \p_a \CJ(x) .
\end{split}
\end{equation}
From \eqref{soft_thm} and \eqref{Ja_def}, it is clear that when inserted into an $S$-matrix element, $N_a(x)$ satisfies the constraint
\begin{equation}
\begin{split}\label{Na_def_1}
\p_{[a} N_{b]}(x) = 0  \quad \implies \quad N_a(x) = \p_a \phi(x)  .
\end{split}
\end{equation}
A more careful derivation of this constraint by demanding the invertibility of the symplectic form was given in \cite{He:2020ifr}.

\subsubsection{Virtual Soft Photons}

Scattering amplitudes in four-dimensional gauge theories formally vanish due to IR divergences. In the perturbative expansion, these arise from diagrams involving exchanges of virtual photons. Each diagram is separately divergent, but the infinite sum exponentiates, and the full amplitude vanishes. Introducing an IR cutoff $\mu$ to regulate the divergences, one finds that an $n$-point amplitude has the form (see Chapter 13 of \cite{Weinberg:1995mt} for details)
\begin{equation}
\begin{split}
\label{ir_factorization}
A_n = e^{-\G(\mu,\L)} {\tilde A}_n ,
\end{split}
\end{equation}
where $\G(\mu,\L)$ captures the IR divergences, $\L$ is the energy scale demarcating the soft from the hard modes, i.e., $\mu \ll \L \ll |p_k^0|$ for all $k$, and ${\tilde A}_n$ is an IR finite amplitude. In abelian gauge theories, the explicit form of $\G$ can be easily worked out to be\footnote{The $i\e$ prescription is a bit different when $k=k'$, in which case $p_{k'} \cdot \ell + i \e$ is replaced with $p_k \cdot \ell - i \e$.} 
\begin{equation}
\begin{split}
\label{Gamma_def}
\G(\mu,\L) = - \frac{ie^2}{2} \sum_{k,k'=1}^n Q_k Q_{k'} \int_\mu^\L \frac{\dt^{d+2} \ell}{(2\pi)^{d+2}} \frac{ p_k \cdot p_{k'}  }{ ( \ell^2 - i \e ) ( p_k \cdot \ell - i \e ) ( p_{k'} \cdot \ell + i \e ) } ,
\end{split}
\end{equation}
where the integration limits above denote integration over the regime $\mu < |\o| < \L$. The explicit form of $\G$ was determined in \cite{Kapec:2021eug} to be\footnote{Actually, $\G$ also has an imaginary part that was calculated in \cite{Kapec:2021eug}, but the soft effective action constructed there (and the one we discuss here) does not address this piece. Hence, we will also only focus on the real part here, and leave the imaginary part of $\G$ to be discussed in future work.}
\begin{equation}
\begin{split}
\label{alpha_def}
\G = \a \int_{\mss^d} \frac{\dt^d x}{(2\pi)^d} \big( \CJ_a(x) \big)^2 , \qquad \a  = \frac{e^2}{8\pi} \int_\mu^\L \dt \o \, \o^{d-3} . 
\end{split}
\end{equation}
We can then write \eqref{ir_factorization} as
\begin{equation}
\begin{split}
\avg{ \CO_1 \cdots \CO_n  }_\mu = \exp \left[ - \a \int_{\mss^d} \frac{\dt^d x}{(2\pi)^d} \big( \CJ_a(x) \big)^2 \right] \avg{ \CO_1 \cdots \CO_n  }_\L  . 
\end{split}
\end{equation}
Note from \eqref{alpha_def} that $\a \to \infty$ as we remove the IR cutoff by taking $\mu \to 0$ in four dimensions ($d=2$), from which we find that $A_n \to 0$. On the other hand, there are no IR divergences in dimensions greater than four ($d>2$) since $\a$ remains finite as $\mu \to 0$, and so amplitudes $A_n$ are not automatically vanishing as $\mu \to 0$.

\subsection{Soft Effective Action}
\label{sec:sea}

Celestial holography postulates that scattering amplitudes in $d+2$ dimensions are correlation functions in a putative holographic conformal field theory in $d$ dimensions. While there have been a few attempts at constructing explicit examples of flat holography \cite{Costello:2022wso, Costello:2022jpg, Costello:2023hmi}, these are only applicable to a very special class of four-dimensional theories. A small step towards a formulation of flat holography in general dimensions was taken in \cite{Kapec:2021eug}, which used effective field theory techniques to construct a $d$-dimensional action that partially reproduces the soft factorization described in the previous section (see also the related works \cite{Nande:2017dba, Kapec:2017gsg, Himwich:2020rro, Arkani-Hamed:2020gyp, Magnea:2021fvy, Gonzalez:2021dxw, Nguyen:2021ydb, Kalyanapuram:2021tnl}). Essentially, the analysis in \cite{Kapec:2021eug} began with the path integral definition of a generic scattering amplitude with $m$ soft particles and $n$ hard particles, given by
\begin{equation}
\begin{split}
& \avg{ N_1 \cdots N_m \CO_1 \cdots \CO_n  }_\mu = \int_\mu [ \dt \varphi ] \, e^{i S[\varphi]} N_1(\varphi) \cdots N_m(\varphi) \CO_1(\varphi) \cdots \CO_n(\varphi) ,
\end{split}
\end{equation}
where the subscript $\mu$ indicates the path integral is over all fields $\varphi$ with $|\o| > \mu$. We can now separate $\varphi$ into a hard piece $\varphi_h$ and a soft piece $\varphi_s$, which respectively have support in the momentum range $|\o| > \L$ and $\mu<|\o|<\L$. By definition, the soft operators depend only on the soft fields, so that $N_i \equiv N_i ( \varphi_s )$, whereas the hard operators factorize as $\CO_k(\varphi) = U_k(\varphi_s) \CO_k(\varphi_h)$ \cite{He:2020ifr}. Substituting these results into the soft theorem \eqref{soft_thm_1} and recalling \eqref{ir_factorization}, we recover the soft factorization
\begin{align}
    A_{m+n} \xrightarrow{q_i \to 0} e^{-\G(\mu,\L)} \SS_m^\0  \tilde A_n ,
\end{align}
with
\begin{equation}
\label{soft-and-hard}
\begin{split}
    {\tilde A}_n &= \avg{ \CO_1 \cdots \CO_n  }_\L = \int_{\L} [ \dt \varphi ] \, e^{i S[\varphi_h]} \CO_1(\varphi_h) \cdots \CO_n(\varphi_h)  \\
\end{split}
\end{equation}
and
\begin{equation}\label{soft-only}
\begin{split}
    e^{-\G(\mu,\L)} \SS^\0_m  &= \avg{ N_1 \cdots N_m U_1 \cdots U_n }_\mu \\
    &= \int_{\mu} [ \dt \varphi_s ] \, e^{- S_{\soft}[\varphi_s]} N_1(\varphi_s) \cdots N_m(\varphi_s) U_1(\varphi_s) \cdots U_n(\varphi_s)  ,
\end{split}
\end{equation}
where $S_{\soft}[\varphi_s]$ is the effective action for the soft modes. This can be constructed by integrating out the hard modes explicitly. However, given the universal IR features that this action is supposed to reproduce, we expect $S_{\soft}[\varphi_s]$ to be universal in any abelian gauge theory. Motivated by this, the authors of \cite{Kapec:2021eug} used general effective field theory ideas to construct the action.

The relevant soft fields in gauge theories are the soft photon operators $N^\pm_a(x)$, defined in \eqref{Napm_def}, and the Goldstone mode for large gauge transformations\footnote{Note that the gauge field satisfies an antipodal matching condition $A_a|_{\CI^+_-} = A_a|_{\CI^-_+}$, so \eqref{Ca_def} could also have been defined on $\CI^-_+$.\label{footnote:match-cond}}
\begin{equation}
\begin{split}
\label{Ca_def}
C_a(x) \equiv A_a |_{\CI^+_-}(x) = \p_a \t(x) , \qquad \t(x) \sim \t(x) + 2\pi . 
\end{split}
\end{equation}
Substituting \eqref{Gamma_def} into \eqref{soft-and-hard}, it was shown in \cite{Kapec:2021eug} that the effective action for the soft modes is given by 
\begin{equation}
\begin{split}
\label{soft-action}
S_{\soft} [\phi,\t] = \a \int_{\mss^d} \frac{\dt^d x}{(2\pi)^d} \big( N_a(x) \big)^2 - \frac{i}{2c_{1,1}} \int_{\mss^d} \dt^d x \,{\wt C}^a(x) N_a(x) ,
\end{split}
\end{equation}
where ${\wt C}_a(x)$ is the shadow transform of $C_a(x)$. For a vector field of scaling dimension $\D$, this is defined by
\begin{equation}
\begin{split}\label{shadow_def}
{\wt C}_a (x) \equiv \int_{\mss^d} \dt^d y \frac{\CI_{ab}(x-y) }{ [ ( x - y )^2 ]^{d-\D} } C^b(y) , \qquad \CI_{ab}(x) \equiv \d_{ab} - 2 \frac{x_a x_b }{x^2}.
\end{split}
\end{equation}
Notice that up to a normalization constant, the shadow transform is its inverse:
\begin{equation}
\begin{split}\label{shadow_sq}
\wt{\wt V}_a(x) = c_{\D,1} V_a(x) , \qquad c_{\D,1} = \frac{\pi^d(\D-1)(d-\D-1) \G(\frac{d}{2}-\D)\G(\D-\frac{d}{2})}{\G(\D+1)\G(d-\D+1)}  . 
\end{split}
\end{equation}
In our case of interest, $C_a(x)$ has scaling dimension $\D=1$, and its shadow transform is evaluated first using \eqref{shadow_def} for generic $\D$ and then taking the limit $\D \to 1$. 

Lastly, the operators $U_k$ are given by
\begin{equation}
\begin{split}
U_k(\t)  =  \exp \left[ i Q_k \int_{\mss^d} \dt^d x \, \t(x) \CK_d ( z_k , x_k ; x ) \right] , \qquad z_k \equiv \frac{m_k}{|\o_k|} , 
\end{split}
\end{equation}
where we have used the momentum parametrization \eqref{mompar}, and $\CK_\D$ is the bulk-to-boundary propagator in Euclidean $\ads_{d+1}$, given by
\begin{equation}
\begin{split}
\label{bulk-to-bdy-prop}
\CK_\D(z , x ; y ) = \frac{\G(\D)}{\pi^{\frac{d}{2}} \G\big(\D-\frac{d}{2}\big)} \left[ \frac{z}{ ( x - y )^2 + z^2} \right]^\D .
\end{split}
\end{equation}
The total product of the operators $U_k$ can be written in a nicer form as
\begin{equation}
\begin{split}
\label{Ui_op}
    e^{-S_\inter[\theta]} &\equiv U_1 ( \t )  \cdots U_n ( \t ) \\
    &= \exp \left[ i  \int_{\mss^d} \dt^d x \, \t(x) \sum_{k=1}^n Q_k \CK_d ( z_k , x_k ; x ) \right] \\
    &= \exp \left[ - \frac{i}{2c_{1,1}} \int_{\mss^d} \dt^d x \, {\wt C}^a(x) \CJ_a(x) \right] ,
\end{split}
\end{equation}
where we recall $\CJ_a(x)$ is the soft factor defined in \eqref{Ja_def}, and in the last equality we used the property
\begin{equation}
\begin{split}\label{Kd_property}
    \sum_{k=1}^n Q_k \CK_d ( z_k , x_k ; x )  = \frac{1}{2 c_{1,1} } \p^a \wt{\CJ}_a(x) ,
\end{split}
\end{equation}
which was derived in \cite{Kapec:2021eug}, as well as the shadow identity
\begin{equation}
\begin{split}
\label{shadow-identity}
    \int_{\mss^d} \dt^d x \, C^a(x) {\wt \CJ}_a(x) = \int_{\mss^d} \dt^d x \, {\wt C}^a(x) \CJ_a(x) . 
\end{split}
\end{equation}
The full soft effective action is then
\begin{align}\label{soft-eff-final}
\begin{split}
    S_\eff[\phi,\theta] &= S_\soft[\phi,\theta] + S_\inter[\theta] \\
    &= \a \int_{\mss^d} \frac{\dt^d x}{(2\pi)^d} \big( N_a(x) \big)^2 - \frac{i}{2c_{1,1}} \int_{\mss^d} \dt^d x \, {\wt C}^a(x) \big( N_a(x) - \CJ_a(x) \big) .
\end{split}
\end{align}

\section{Soft On-Shell Action \texorpdfstring{$\to$}{to} Soft Effective Action}\label{sec:edge-mode}

In this section, we show that the soft effective action \eqref{soft-eff-final} can be obtained by evaluating the bulk gauge theory action on-shell given a specific choice of boundary conditions, and then extracting the contribution from the soft modes. Consider a generic model describing an abelian gauge field coupled to charged matter, which is described by a Lagrangian of the form
\begin{equation}
\begin{split}
S = \int_\CM \left(  - \frac{1}{2e^2} F \w \star F + L_M \right)  + S_{\text{bdy}},
\end{split}
\end{equation}
where $L_M$ is the rest of the Lagrangian and includes all the matter field contributions and any potential higher derivative terms in the Lagrangian. Generically, it is a polynomial function of the arguments
\begin{equation}
\begin{split}
L_M \equiv L_M \bigg( \p_{A_1} \cdots \p_{A_n} F_{AB} , D_{(A_1} \cdots D_{A_n)} \Phi^i \bigg) , \qquad D_A \Phi^i \equiv \p_A - i Q_i A_A \Phi^i ,
\end{split}
\end{equation}
where $D_{(A_1} \cdots D_{A_n)}$ denotes $n$ symmetrized covariant derivatives.\footnote{The commutator of covariant derivatives simplifies to the field strength, in that $[ D_A , D_B ] \Phi^i = - i Q_i F_{AB} \Phi^i$. Thus, without loss of generality, it suffices to consider symmetrized derivatives.} In particular, we are interested in the leading order contribution of the soft gauge field modes to the on-shell action, which would arise from the lowest derivative terms in the action. Now, after integrating out the matter fields, what remains at the lowest derivative order is a term of the form $A_A J^A$, where $J_A$ is a background conserved current that is determined from the boundary conditions used for the charged matter fields. To summarize, as far as the contribution of the soft modes is concerned, we can restrict ourselves to a simple model described by the action
\begin{equation}
\begin{split}
\label{S_action}
S[A] = \int_\CM \left(  - \frac{1}{2e^2} F \w \star F + (-1)^d A \w \star J \right)   + \frac{1}{e^2} \int_{\S^+} A \w \star F - \frac{1}{e^2} \int_{\S^-} A \w \star F ,
\end{split}
\end{equation}
where $\S^\pm = \CI^\pm \cup i^\pm$. A model of this type was considered in \cite{Hirai:2020kzx, Delisle:2020uui, Hirai:2021ddd}, where it was indeed shown to reproduce all the IR effects described earlier in Section~\ref{ssec:soft-factor}. 

Let us begin by focusing on the boundary terms in \eqref{S_action}, which are required so that the variational principle imposes the relevant boundary conditions for our model. To see this, note that the variation of the action has the form 
\begin{equation}
\begin{split}
\label{S_action_1}
\d S[A] &=  - \frac{1}{e^2} \int_\CM  \d A \w \left( \dt \star F - (-1)^d e^2 \star J \right) \\
&\qquad + \frac{1}{e^2} \int_{\S^+} A \w \star \d F  - \frac{1}{e^2} \int_{\S^-} A \w \star \d F  - \frac{1}{e^2} \int_{i^0} \d A \w \star F . 
\end{split}
\end{equation}
The first term in \eqref{S_action_1} gives us Maxwell's equations\footnote{Notice that \eqref{maxwell_eq} also implies current conservation, since acting on both sides by $\dt$ yields $\dt \star J = 0$, or $\p^A J_A = 0$.}
\begin{equation}
\begin{split}
\label{maxwell_eq}
\dt \star F = (-1)^d e^2 \star J \quad \implies \quad \p^A F_{AB}(X) = e^2 J_B(X) .
\end{split}
\end{equation}
Furthermore, the variational principle holds only if the terms in the second line of \eqref{S_action_1} vanishes, which requires us to impose Neumann boundary conditions on $\S^\pm$ and Dirichlet boundary conditions on $i^0$, so that
\begin{equation}
\begin{split}
\label{bdy_cond}
\d A |_{i^0} = 0  ,\qquad \iota_n \d F  |_{\S^\pm} = 0 ,
\end{split}
\end{equation}
where $\iota_n$ is the interior product with respect to the normal vector $n^A$, i.e., $(\iota_n F)_A = n^B F_{BA}$.

Before continuing, we note that for the calculations presented in this section, we did not need to know the precise form of the background current $J_A$ (aside from the fact that it is conserved). However, to match the results here to those of Sections~\ref{ssec:soft-factor} and \ref{sec:sea}, we will need the current to be the one corresponding to $n$ charged point-particles (this is the relevant choice for the scattering problem), so that
\begin{equation}
\begin{split}
\label{JA-explicit}
J^A(X) &= - \sum_{k=1}^n  \t(\eta_k X^0) \eta_k Q_k \frac{ p_k^A}{p^0_k} \d^{(d+1)} \left( \vec{X}  - \frac{\vec{p}_k}{p^0_k} X^0 \right) ,
\end{split}
\end{equation}
where $\eta_k=\pm1$ distinguishes outgoing ($+$) particles from incoming ($-$) ones.\footnote{Notice that \eqref{JA-explicit} assumes that the scattering takes place at a single point $X^A = 0$. This is of course not true for a generic scattering process, but since we are only interested in the leading soft (IR) behavior of the current, the actual details of the scattering process are not relevant, and \eqref{JA-explicit} is a reasonable approximation.}

\subsection{Solutions to Maxwell's Equations}
\label{solution}

Since we aim to evaluate the action on-shell, we start by discussing solutions to \eqref{maxwell_eq}. We work in axial null gauge, given by\footnote{Notice that $A_u =  n^A A_A$, so the axial null gauge \eqref{null_gauge} is the same as imposing $A_u = 0$, which was previously called temporal gauge in \cite{He:2020ifr, He:2023bvv}.\label{fn:gauge-condition}}
\begin{equation}
\begin{split}
\label{null_gauge}
n^A A_A(X) = 0 . 
\end{split}
\end{equation}
To solve \eqref{maxwell_eq}, we decompose the gauge field into the pieces
\begin{equation}
\begin{split}
\label{A_decomposition}
A_A(X) = {\hat A}_A(X) + \p_A \t(X) , \qquad \t(X) \sim \t(X) + 2\pi ,
\end{split}
\end{equation}
where $\t(X)$ captures the Goldstone mode for large gauge transformations, and ${\hat A}_A(X)$ is the part of the gauge field that admits a Fourier transform, namely
\begin{equation}
\begin{split}
\label{Ahat_Fourier}
{\hat A}_A(X) = e \int_\CM \frac{\dt^{d+2}\ell}{(2\pi)^{d+2}} e^{i \ell \cdot X} {\hat A}_A(\ell) . 
\end{split}
\end{equation}
Similarly, we consider the Fourier transform the current, given by
\begin{equation}
\begin{split}
J_A(X) =  \int_\CM \frac{\dt^{d+2}\ell}{(2\pi)^{d+2}} e^{i \ell \cdot X} J_A(\ell) , \qquad \ell^A J_A(\ell) = 0 ,
\end{split}
\end{equation}
where the second equality is due to current conservation $\p^A J_A(X)  = 0$. Substituting the Fourier modes into \eqref{maxwell_eq}, we obtain
\begin{equation}
\begin{split}
\label{Maxwell_Eq_Fourier}
\ell^2 {\hat A}_A(\ell) - \ell_A \ell^B {\hat A}_B(\ell) = - e J_A(\ell).
\end{split}
\end{equation}
To solve this, we will find it convenient to expand the gauge field and current using the basis of vectors $\{n^A, \ell^A , \ve^A_a(\ell) \}$ on $\mrr^{1,d+1}$, where the polarization vectors $\ve^A_a(\ell)$ are defined in \eqref{polpar} (with $x^a$ related to $\ell^A$ via \eqref{offshell_mom}) and satisfy the properties
\begin{equation}
\begin{split}
\label{pol_properties}
n_A \ve^A_a(\ell) = \ell_A \ve^A_a(\ell) = 0 , \qquad \eta_{AB} \ve^A_a(\ell) \ve^B_b(\ell) = \d_{ab} , \qquad \ve^A_a(\ell) = \ve^A_a(-\ell) = \ve^A_a(\ell)^* .
\end{split}
\end{equation}
Expanding the gauge field and current in this basis, we find
\begin{equation}
\begin{split}
{\hat A}_A(\ell) = n_A L(\ell) + \ve_A^a(\ell) {\hat A}_a(\ell) , \qquad J_A(\ell) = \left( \frac{\ell_A}{n \cdot \ell} - \frac{\ell^2 n_A}{(n \cdot \ell )^2} \right) J_n(\ell)  + \ve_A^a(\ell) J_a(\ell) ,
\end{split}
\end{equation}
where the coefficients are fixed by imposing the gauge condition \eqref{null_gauge} and current conservation $\ell^A J_A(\ell) = 0$. Substituting this result into \eqref{Maxwell_Eq_Fourier}, we obtain
\begin{equation}
\begin{split}
L(\ell)  = e \frac{J_n(\ell)}{(n\cdot \ell)^2}  , \qquad \ell^2 {\hat A}_a(\ell) =  - e J_a(\ell)  . 
\end{split}
\end{equation}
The second equation above solves to
\begin{equation}
\begin{split}
\label{Ahat_sol}
{\hat A}_a(\ell) = 2\pi \CO^\text{rad}_a(\ell) \d(\ell^2) - \frac{eJ_a(\ell)}{\ell^2} ,
\end{split}
\end{equation}
where the first term $\CO^\text{rad}_a(\ell)$ is the homogeneous (radiative) solution, and the second term is the Coulombic solution. We would now like to substitute this result into \eqref{Ahat_Fourier} to determine the gauge field in position space. However, to evaluate this Fourier integral, the pole at $\ell^2=0$ in the second term of \eqref{Ahat_sol} has to be regulated by an $i\e$ prescription. Depending on how this is done, the corresponding radiative solution is incoming or outgoing. More precisely, we have
\begin{equation}
\begin{split}
\label{Ahat_sol_reg}
{\hat A}_a(\ell) = 2\pi \CO^\pm_a(\ell) \d(\ell^2) - \frac{e J_a(\ell)}{-(\ell^0 \mp i \e )^2 + | \vec{\ell}\, |^2 } ,
\end{split}
\end{equation}
where, as before, the $\pm$ superscript corresponds to the outgoing ($+$) and incoming ($-$) radiative modes, respectively. Furthermore, depending on the sign of $\ell^0=\pm|\vec{\ell}\,|$, the operator $\CO_a^\pm(\ell)$ reduces to a creation or an annihilation operator in the quantum theory, and we have the identification
\begin{equation}
\begin{split}
\CO^\pm_a(|\vec{\ell}\,|,\vec{\ell}\,)  = \CO^\pm_a(\vec{\ell}\,) , \qquad \CO^\pm_a(-|\vec{\ell}\,|,-\vec{\ell}\,)  = \CO^{\pm}_a(\vec{\ell}\,)^\dag . 
\end{split}
\end{equation}
Finally, we remark that using the identity
\begin{equation}\label{delta-id}
\begin{split}
    \frac{1}{x \pm i\e} = \CP \left( \frac{1}{x} \right) \mp i \pi \d ( x ) , 
\end{split}
\end{equation}
where $\CP$ is the Cauchy principal value, a useful consequence of \eqref{Ahat_sol_reg} is
\begin{equation}
\begin{split}
\label{res_1}
\left[ J_a(\ell) - \frac{i}{e} \T(\ell^0) \big(\CO^+_a(\ell) -  \CO^-_a(\ell) \big)  \right] \d ( \ell^2 ) = 0 ,
\end{split}
\end{equation}
where $\Th$ is the sign function.

\subsection{On-Shell Action}

Having constructed the solutions, we now turn to the on-shell action. First, using the decomposition \eqref{A_decomposition}, the action \eqref{S_action} can be recast into the form 
\begin{equation}
\begin{split}
\label{Action_Expanded}
    S[A] &= \int_\CM \left(   - \frac{1}{2e^2} {\hat F} \w \star {\hat F} + (-1)^d {\hat A} \w \star J \right) - \frac{1}{e^2} \int_{\CI_-^+} \t \star {\hat F} - \frac{1}{e^2} \int_{\CI_+^-} \t \star {\hat F} \\
    &\qquad  + \frac{1}{e^2} \int_{\S^+} {\hat A} \w \star \hat F - \frac{1}{e^2} \int_{\S^-} {\hat A} \w \star \hat F + (-1)^d \int_{i^0} \t \star J   \\
    &\qquad  - \frac{1}{e^2} \int_{\S^+}  \t \left(  \dt \star \hat F - (-1)^d e^2 \star J \right)  + \frac{1}{e^2} \int_{\S^-} \t \left(  \dt \star \hat F - (-1)^d e^2 \star J \right)  . 
\end{split}
\end{equation}
The terms in the last line are proportional to the equations of motion \eqref{maxwell_eq} and therefore vanish on-shell. Furthermore, all the terms in the second line vanish on-shell as well. To see why, first note that there is no charge flux through $i^0$, so the last term in the second line vanishes. Secondly, using \eqref{volume_forms}, the first two terms in the second line can be written as\footnote{We remark that although \eqref{eq:term1-inter} only includes the contribution from $\ci^\pm$, we are allowed to drop the contribution from $i^\pm$. This is because in the absence of massive particles, the gauge field vanishes on $i^\pm$, while in the presence of massive particles, the gauge field only receives a Coulombic contribution on $i^\pm$ (see Appendix~\ref{app:massive}), which falls off too quickly to contribute to \eqref{eq:term1-inter}.}
\begin{align}\label{eq:term1-inter}
\begin{split}
    \frac{1}{e^2} \int_{\S^\pm} \hat A \wedge \star \hat F &= -\frac{1}{e^2} \int_\mrr \dt u \int_{\mss^d} \dt^dx \lim_{r \to \pm \infty}  \, |r|^{d} \hat A_a \p_u \hat A^a .
\end{split}    
\end{align}
Following \cite{He:2023bvv}, we now decompose the gauge field into radiative and Colulombic modes, so that 
\begin{align}
	\hat A_\mu = \hat A_\mu^{R \pm} + \hat A_\mu^{C \pm},
\end{align}
where the radiative piece ($R$) is the homogeneous solution to Maxwell's equations, and the Coluombic piece ($C$) is the inhomogenous solution. The $\pm$ superscript indicates whether we are taking an advanced Green's function ($+$) or retarded Green's function ($-$). The fall-off conditions for these pieces obey \cite{He:2023bvv}
\begin{align}\label{falloff}
\begin{split}
	\hat A_r^{R\pm} &= \CO(|r|^{-\frac{d}{2}-1}) + \CO(|r|^{-d}) , \qquad \hat A_r^{C \pm} = \CO(|r|^{-d}) , \\
	\hat A_a^{R\pm} &= \CO(|r|^{-\frac{d}{2}+1}) + \CO(|r|^{-d+1}) , \qquad \hat A_r^{C \pm} = \CO(|r|^{-d+1}) .
\end{split}
\end{align}
Given these fall-off conditions, it is clear from the integrand in \eqref{eq:term1-inter} that only the radiative part of the gauge field contributes to the integral, as the Coulombic modes fall off too quickly as $|r| \to \infty$. It follows we have
\begin{align}\label{vanish-inter}
\begin{split}
	\frac{1}{e^2} \int_{\S^\pm} \hat A  \wedge \star \hat F = -\frac{1}{e^2} \int_\mrr \dt u \int_{\mss^d} \dt^d x \, \hat A_a^\pm \p_u \hat A^{\pm a},
\end{split}
\end{align}
where 
\begin{align}\
	\hat A_a^\pm(u, x) = \lim_{r \to \pm\infty} |r|^{\frac{d}{2}-1} \hat A_a^{\pm}(u,r,x) .
\end{align}
As $\hat A_a^\pm$ only involves the radiative modes, it admits a mode expansion, which is given on-shell in \cite{He:2023bvv} to be
\begin{align}
\begin{split}
	\hat A_a^\pm(u,x) &=  \pm \frac{e}{2(2\pi)^{\frac{d}{2}+1}} \int_0^\infty \dt \o\, \o^{\frac{d}{2}-1} \Big[ \CO^\pm_a(\o\hat q(x)) e^{-\frac{i\o u}{2} \mp \frac{i\pi d}{4}} + \cc \Big] .
\end{split}
\end{align}
Substituting this into \eqref{vanish-inter}, we obtain
\begin{align}\label{vanish-final}
\begin{split}
	\frac{1}{e^2} \int_{\S^\pm} \hat A \wedge \star \hat F &= -\frac{i}{8(2\pi)^{d+2}} \int_\mrr \dt u \int_{\mss^d} \dt^d x\,  \int_0^\infty \dt \o \, \dt \o' \, \o^{\frac{d}{2}-1} \o'^{\frac{d}{2}}   \\
	&\qquad \qquad \Big[ \CO^\pm_a(\o \hat q(x) ) e^{-\frac{i\o u}{2} \mp \frac{i\pi d}{4}} + \cc \Big] \Big[ \CO^{\pm a}(\o' \hat q(x) ) e^{-\frac{i\o' u}{2} \mp \frac{i\pi d}{4}} - \cc \Big]  \\
	&= \frac{i}{8(2\pi)^{d+2}}  \int_{\mss^d} \dt^d x\,  \int_0^\infty \dt \o \, \dt \o' \, \o^{\frac{d}{2}-1} \o'^{\frac{d}{2}} \\
 &\qquad \qquad \times \int_\mrr \dt u\, \Big[ \CO_a^\pm(\o \hat q(x)) \CO^{\pm a}(\o'\hat q(x))^\dag e^{-\frac{i(\o-\o') u}{2} } - \cc \Big] \\
	&= 0 .
\end{split}
\end{align}
This proves the claim that all the terms in the second line of \eqref{Action_Expanded} vanish.

Thus, we see that on-shell, the only terms that survive in \eqref{Action_Expanded} are those in the first line, i.e.,
\begin{align}\label{Action_Expanded2}
S[A] \big|_{\text{on-shell}} &= \int_\CM \left(   - \frac{1}{2e^2} {\hat F} \w \star {\hat F} + (-1)^d {\hat A} \w \star J \right) - \frac{1}{e^2} \int_{\CI_-^+} \t \star {\hat F} - \frac{1}{e^2} \int_{\CI_+^-} \t \star {\hat F} .
\end{align}
In the rest of this subsection, we will demonstrate that with a suitable contour deformation, the bulk integral in \eqref{Action_Expanded2} in the soft and on-shell limit becomes (see \eqref{bulk-result})
\begin{align}\label{bulk-claim}
\begin{split}
    \int_\CM \left(- \frac{1}{2e^2} {\hat F} \w \star {\hat F} + (-1)^d {\hat A} \w \star J \right) &\quad \xrightarrow{\text{soft}+\text{on-shell}}\quad   i \a \int_{\mss^d} \frac{\dt^d x}{(2\pi)^d} \big( N_a(x) \big)^2 ,
\end{split}
\end{align}
and the boundary integrals in \eqref{Action_Expanded2} in the on-shell limit becomes (see \eqref{eq:codim2-result})\footnote{Only the soft modes are non-vanishing on $\CI^\pm_\mp$, so we do not need to take a soft limit.}
\begin{align}\label{bdy-claim}
    - \frac{1}{e^2} \int_{\CI_-^+} \t \star {\hat F} - \frac{1}{e^2} \int_{\CI_+^-} \t \star {\hat F}  &\quad \xrightarrow{\text{on-shell}}\quad \frac{1}{2c_{1,1}} \int_{\mss^d} \dt^d x \, \wt{C}^a(x) \big( N_a(x) - \CJ_a(x) \big) . 
\end{align}
where $C_a(x)$ was defined in \eqref{Ca_def}. Substituting \eqref{bulk-claim} and \eqref{bdy-claim} into \eqref{Action_Expanded2}, we see that the soft limit of the on-shell action (with a suitable contour deformation) is given by
\begin{align}\label{on-shell-final}
    S[A]\big|_{\soft \text{+on-shell}} = i \a \int_{\mss^d} \frac{\dt^d x}{(2\pi)^d} \big( N_a(x) \big)^2 + \frac{1}{2c_{1,1}} \int_{\mss^d} \dt^d x \, \wt{C}^a(x) \big( N_a(x) - \CJ_a(x) \big) . 
\end{align}
Comparing with the soft effective action \eqref{soft-eff-final}, we see that 
\begin{align}\label{final-result}
\begin{split}
    S[A]\big|_{\soft \text{+on-shell}} = i S_\eff[\phi,\theta] .
\end{split}
\end{align}
This is the main result of our paper, and it proves our claim that the soft limit of on-shell degrees of freedom localized on the celestial sphere, i.e., the soft limit of edge modes, are precisely the soft and Goldstone modes parametrizing the low-energy Hilbert space of the gauge theory. To understand the factor of $i$, note that from \eqref{soft-only}, the path integral involves $e^{-S_\eff[\phi,\theta]}$. On the other hand, if we had chosen instead to insert the on-shell action into the path integral, it would involve $e^{i S[A]|_\text{on-shell}}$, implying \eqref{final-result} is indeed correct. In the next two subsubsections, we will derive both \eqref{bulk-claim} and \eqref{bdy-claim}, which were necessary to prove the main result \eqref{final-result}.

\subsubsection{Bulk Term}

We start with the first term in \eqref{Action_Expanded2}, which in momentum space can be written as
\begin{equation}
\begin{split}
S_{\bulk}[{\hat A}] =  \int_\CM \frac{\dt^{d+2}\ell}{(2\pi)^{d+2}}  \left[  - \frac{1}{2} \left(  \ell^2  {\hat A}(\ell) \cdot {\hat A}(-\ell)  - | \ell \cdot {\hat A}(\ell) |^2 \right) - e  {\hat A}(\ell) \cdot J(-\ell) \right] . 
\end{split}
\end{equation}
To render the Lorentzian path integral finite, we need to deform the contour of integration over $\ell$. A simple way to do this is to replace $\ell^2 \to \ell^2 - i \e$ above. Applying this deformation and then evaluating the action on-shell by utilizing the solutions constructed in Section~\ref{solution}, we find
\begin{equation}
\begin{split}
S_{\bulk}[{\hat A}]\big|_{\text{on-shell}} = \frac{e^2}{2} \int_\CM \frac{\dt^{d+2}\ell}{(2\pi)^{d+2}} \frac{1}{\ell^2-i\e} \left[ |J_a(\ell)|^2 - \frac{\ell^2}{(n \cdot \ell)^2}   | J_n(\ell)|^2   \right]  . 
\end{split}
\end{equation}
Next, using the identity \eqref{delta-id} and \eqref{res_1}, we can rewrite the action as
\begin{equation}
\begin{split}
\label{S1_os}
S_{\bulk}[{\hat A}]\big|_{\text{on-shell}} &= \int_\CM \frac{\dt^{d+2}\ell}{(2\pi)^{d+2}} \left[ \frac{i \pi}{2} \d(\ell^2)  \big| \CO^+_a(\ell) -  \CO^-_a(\ell) \big|^2 + \frac{e^2}{2}  \left( \frac{\CP(|J_a(\ell)|^2)}{\ell^2} - \frac{|J_n(\ell)|^2}{(n \cdot \ell)^2}  \right) \right] . 
\end{split}
\end{equation}
For the soft effective action, we only keep the first term above, as this is the term that is responsible for the real part of the IR divergence $\G$.\footnote{The imaginary part of $\G$ is related to the second term of \eqref{S1_os}, and as mentioned previously was ignored in \cite{Kapec:2021eug}. Since our goal in this paper is to reproduce the results of \cite{Kapec:2021eug}, we will ignore the second term in \eqref{S1_os} for now, and hope to return to it in future work.} Extracting the soft contribution here, we find
\begin{equation}
\begin{split}
    S_{\bulk}[{\hat A}]\big|_{\text{on-shell+soft}} &= \frac{i \pi}{2} \int_\mu^\L \frac{\dt^{d+2}\ell}{(2\pi)^{d+2}} \d(\ell^2)  \big| \CO^+_a(\ell) -  \CO^-_a(\ell) \big|^2 .
\end{split}
\end{equation}
Using \eqref{offshell_int}, this can be rewritten as
\begin{equation}
\begin{split}
    S_{\bulk}[{\hat A}]\big|_{\text{on-shell+soft}} &= \frac{i}{8\pi} \int_\mu^\L \dt \o \,\o^{d-1} \int_{\mss^d} \frac{\dt^d x}{(2\pi)^d}    \big| \CO^+_a(\o {\hat q}(x)) - \CO^-_a(\o {\hat q}(x)) \big|^2 .
\end{split}
\end{equation}
We now recall that $\mu$ and $\L$ are much smaller than any other scale in the problem, so using \eqref{Napm_def}, we can write
\begin{equation}
\begin{split}
\CO^\pm_a(\o {\hat q}(x)) \to \frac{e}{\o} N_a^\pm(x) ,
\end{split}
\end{equation}
which implies
\begin{equation}\label{bulk-result}
\begin{split}
    S_{\bulk}[{\hat A}]\big|_{\text{on-shell+soft}} &= i \a \int_{\mss^d} \frac{\dt^d x}{(2\pi)^d} \big( N_a(x) \big)^2 ,
\end{split}
\end{equation}
where $N_a$ is defined in \eqref{Na_def} and $\a$ in \eqref{alpha_def}. This proves our claim \eqref{bulk-claim}.

\subsubsection{Boundary Terms}
\label{sec:boundary_terms}

We now turn to the boundary terms in \eqref{Action_Expanded2}, which are 
\begin{align}
\label{Sbdy}
S_{\bdy}[\hat A,\theta] = -\frac{1}{e^2} \int_{\ci^+_-} \theta \star \hat F - \frac{1}{e^2} \int_{\ci^-_+} \theta \star \hat F.
\end{align}
Using \eqref{volume_forms}, we can write the terms in coordinate notation as
\begin{align}
\label{term2}
\begin{split}
\frac{1}{e^2} \int_{\ci^\pm_\mp} \theta  \star \hat F &= \frac{2}{e^2} \int_{\mss^d}  \dt^d x \, \t(x)  \left( \lim_{u \to \mp\infty} \lim_{r \to \pm \infty} |r|^d \hat F_{ur} (u,r,x) \right) ,
\end{split}
\end{align}
where we have used the matching condition on the gauge field $\t|_{\CI^+_-} = \t|_{\CI^-_+}$ (see Footnote \ref{footnote:match-cond}). To evaluate \eqref{term2} explicitly, we decompose the field strength into radiative and Coulombic parts, so that
\begin{align}
\begin{split}
{\hat F}_{ur} (u,r,x)  = {\hat F}_{ur}^{R\pm}  (u,r,x)  + {\hat F}_{ur}^{C\pm}  (u,r,x)  . 
\end{split}
\end{align}
From \cite{He:2023bvv}, we have for abelian gauge theories the \emph{on-shell} identity
\begin{align}\label{R-useful}
\begin{split}
\left( \lim_{u \to \mp\infty} \lim_{r \to \pm\infty} |r|^d F_{ur}^{R\pm}  (u,r,x) \right) = \pm \frac{e^2}{4c_{1,1}} \p^a \wt{N}^\pm_a (x)  .
\end{split}
\end{align}
Furthermore, Maxwell's equations imply \cite{He:2019pll}
\begin{align}\label{C-useful}
\begin{split}
2 \p_u \left( \lim_{r \to \pm\infty} |r|^d {\hat F}_{ur}^{C\pm}  (u,r,x) \right) &= e^2 J_u^\pm (u,x) , \qquad J_u^\pm(u,x) \equiv \lim_{r \to \pm\infty} |r|^d J_u (u,r,x)  . 
\end{split}
\end{align}
Integrating this differential equation in $u$, we obtain
\begin{equation}
\begin{split}
\label{FurC_result}
    \left(  \lim_{u \to \mp \infty} \lim_{r \to \pm\infty} |r|^d {\hat F}_{ur}^{C\pm}  (u,r,x)  \right) &= \mp \frac{e^2}{2}\int_\mrr \dt u \, J_u^\pm(u,x) + \left(  \lim_{u \to \pm \infty} \lim_{r \to \pm\infty} |r|^d {\hat F}_{ur}^{C\pm}   (u,r,x) \right) .
\end{split}
\end{equation}
In the absence of massive particles (the contribution from massive particles is discussed in Appendix \ref{app:massive}), the second term above vanishes from the fall-off condition \eqref{falloff}. Assuming this and substituting \eqref{R-useful} and \eqref{FurC_result} into \eqref{term2}, we get
\begin{align}
\begin{split}
    \frac{1}{e^2} \int_{\ci^\pm_\mp} \theta  \star \hat F &= \pm  \frac{1}{2c_{1,1}} \int_{\mss^d}  \dt^d x \,  \t(x) \p^a \wt{N}^\pm_a(x) \mp \int_{\mss^d}  \dt^d x \, \t(x)  \int_\mrr \dt u \, J_u^\pm(u, x) \\
    &= \mp \frac{1}{2c_{1,1}} \int_{\mss^d} \dt^d x \, {\wt C}^a(x) N^\pm_a(x) \mp \int_{\mss^d} \dt^d x \, \t(x)  \int_\mrr \dt u \, J_u^\pm(u, x) ,
\end{split}
\end{align}
where in the second equality we integrated by parts in the first term and then used the shadow identity \eqref{shadow-identity}. Substituting this into \eqref{Sbdy}, it follows that
\begin{align}\label{2ndterm2}
\begin{split}
    S_{\bdy}[\hat A,\theta] &= \frac{1}{2c_{1,1}} \int_{\mss^d} \dt^d x \, \wt{C}^a(x) N_a(x) + \int_{\mss^d} \dt^d x \, \theta(x) \int_\mrr \dt u\, \big( J_u^+(u,x)  - J_u^-(u,x) \big) ,
\end{split}
\end{align}
where we used the definition \eqref{Na_def}. To further simplify this expression, we use the following identity, which we prove in Appendix~\ref{app:current-shadow},
\begin{align}\label{identity}
\begin{split}
    \int_\mrr \dt u \,\big(  J_u^+(u,x) - J_u^-(u,x) \big) = \frac{1}{2c_{1,1}} \p^a \wt \CJ_a(x),
\end{split}
\end{align}
where
\begin{align}\label{CJ-def}
\begin{split}
    \CJ_a(x) \equiv \frac{i}{2} \left( \lim_{\o \to 0^+} + \lim_{\o \to 0^-} \right) \o \ve^A_a(x) \int_\CM \dt^{d+2}X\, e^{i \o \hat q(x) \cdot X } J_A(X) ,
\end{split}
\end{align}
and importantly reduces precisely to our earlier definition of $\CJ_a$ in \eqref{Ja_def} when the spacetime current $J_A(X)$ is the point-particle current \eqref{JA-explicit}. Using \eqref{identity}, we can rewrite \eqref{2ndterm2} as
\begin{align}
\label{eq:codim2-result}
\begin{split}
    S_{\bdy}[\hat A,\theta] &= \frac{1}{2c_{1,1}} \int_{\mss^d} \dt^d x \, \wt{C}^a(x) \big( N_a(x) - \CJ_a(x) \big) , 
\end{split}
\end{align}
which proves \eqref{bdy-claim}.

\section{Summary}\label{sec:summary}

We have in this paper shown that the soft effective action \eqref{intro:sea} can be derived from a general action for an abelian gauge theory \eqref{intro:S_action}, taken on-shell in the soft limit, and this result is summarized in \eqref{final-result0}. Importantly, our analysis fixes the \emph{type} of boundary conditions necessary to derive the soft effective action. In particular, the soft modes are \emph{not} the entanglement edge modes studied by Donnelly and Wall in \cite{Donnelly:2014fua, Donnelly:2015hxa}, which analyzed edge modes of Maxwell theory with magnetic conductor boundary conditions imposed. Rather, they are the edge modes for gauge theory with Neumann boundary conditions at timelike and null infinity and Dirichlet boundary conditions at spatial infinity. It would be very interesting to explore the entanglement properties of soft modes by viewing them as entanglement edge modes and following, in spirit, the analysis of Donnelly and Wall. We will leave such explorations for future work.

Furthermore, now that the connection between soft modes and edge modes has been established in abelian gauge theories, there are natural extensions of our analysis to both nonabelian gauge theories and gravity. By beginning with the action in nonabelian gauge theory or gravity with suitable boundary terms added to impose Neumann boundary condition on $\Sigma^\pm$, we can derive the on-shell action. By then taking the soft limit, it would be interesting to confirm that we get precisely the soft effective action for nonabelian gauge theory and gravity given in \cite{Magnea:2021fvy, Kapec:2021eug}.

Indeed, it would be most interesting to study the IR sector of gravity, which is expected to have similar behavior as the IR sector of abelian gauge theories at leading order in small energies. In particular, we would like to determine what are the appropriate GHY boundary terms to add to the Einstein-Hilbert action such that in the soft and on-shell limit it becomes the gravitational soft effective action. This should allow us to appropriately identify soft modes with entanglement edge modes in gravity and gain insight into the modular Hamiltonian. The modular Hamiltonian has been an object of study in connection to quantum fluctuations in spacetime subregions~\cite{Verlinde:2019ade,Verlinde:2019xfb,Verlinde:2022hhs}, and we expect this will open the possibility of utilizing soft or edge modes to study subregion spacetime fluctuations.  These, and many other exciting connections between soft modes and entanglement, are current avenues under exploration.

\section*{Acknowledgements}

We would like to thank Sangmin Choi, Laurent Freidel, Per Kraus, Andrea Puhm, Ana-Maria Raclariu, Antony Speranza, and Andrew Strominger for useful conversations. T.H., A.S., and K.Z. are supported by the Heising-Simons Foundation ``Observational Signatures of Quantum Gravity'' collaboration grant 2021-2817, the
U.S. Department of Energy, Office of Science, Office of High Energy Physics, under Award
No. DE-SC0011632, and the Walter Burke Institute for Theoretical Physics. P.M. is supported
by the European Research Council (ERC) under the European Union’s Horizon 2020 research and
innovation programme (grant agreement No 852386). 
The work of K.Z. is also supported by a Simons Investigator award.

\appendix

\section{Relating Current to its Shadow}
\label{app:current-shadow} 

In this appendix, we will prove the identity \eqref{identity}. Given a conserved current $J_A(X)$, we want to compute the soft limit of its Fourier transform, which we defined symmetrically to be (see \eqref{CJ-def})
\begin{align}
\begin{split}
	\CJ_a(x) &\equiv \frac{i}{2} \left( \lim_{\o \to 0^+} + \lim_{\o \to 0^-} \right) \o \ve^A_a (x) \int_\CM \dt^{d+2}X\, e^{i \o \hat q(x) \cdot X } J_A(X)   .
\end{split}
\end{align}
We now compute using flat null coordinates
\begin{align}\label{eq:Ja-fourier}
\begin{split}
    \int_\CM \dt^{d+2}X\, e^{i \o \hat q(x) \cdot X }  J_A(X) &= \frac{1}{|\o|^{d+1}}\int_\mrr \dt u \int_{\mrr} \dt r \int_{\mss^d} \dt^d y\, \frac{|r|^d}{2} e^{-\frac{i\o}{2} u - \frac{i}{2}r \Th(\o)(x-y)^2} J_A \left(u,\frac{r}{|\o|},y\right),
\end{split}
\end{align}
where we recall $\Th$ is the sign function. Denoting
\begin{align}\label{large-r-J}
	J_A^\pm(u,y) = \lim_{r\to\pm\infty} |r|^d J_A(u,r,y),
\end{align}
we have
\begin{align}
\begin{split}
	\CJ_a(x) &= \frac{i}{2} \bigg( \lim_{\o \to 0^+} + \lim_{\o \to 0^-} \bigg) \p_a \hat q^A(x) \Th(\o) \int_{-\infty}^0 \dt r \int_\mrr \dt u \int_{\mss^d} \dt^dy \,\frac{1}{2} e^{-\frac{i\o}{2} u - \frac{i}{2}r \Th(\o)(x-y)^2}  J_A^- (u, y ) \\
	&\qquad + \frac{i}{2} \bigg( \lim_{\o \to 0^+} + \lim_{\o \to 0^-} \bigg) \p_a \hat q^A(x) \Th(\o) \int_{0}^\infty \dt r \int_\mrr \dt u \int_{\mss^d} \dt^dy \, \frac{1}{2} e^{-\frac{i\o}{2} u - \frac{i}{2}r \Th(\o)(x-y)^2}  J_A^+ (u, y ) ,
\end{split}
\end{align}
where we used \eqref{large-r-J} and the fact the $\o \to 0^\pm$ limit for $J_A$ corresponds to the large $r$ limit. We can now perform the $r$ integral directly, where we have to regulate using the $i\e$ prescription:
\begin{align}
\begin{split}
    \int_{-\infty}^0 \dt r\,e^{-\frac{i}{2} r\Th(\o) \left[ (x-y)^2 + i\Th(\o)\e \right]} &= \frac{e^{-\frac{i}{2} r \Th(\o) \left[ (x-y)^2 + i\Th(\o)\e \right]}}{-\frac{i}{2}\Th(\o)\left[ (x-y)^2 + i\Th(\o)\e \right]} \bigg|_{r=-\infty}^{r = 0}  = \frac{2i\Th(\o)}{ (x-y)^2 + i\Th(\o)\e } \\
    \int_{0}^\infty \dt r\,e^{-\frac{i}{2}r\Th(\o) \left[ (x-y)^2 - i\Th(\o)\e \right]} &= \frac{e^{-\frac{i}{2} r\Th(\o) \left[ (x-y)^2 - i\Th(\o)\e \right]}}{-\frac{i}{2} \Th(\o)\left[ (x-y)^2 - i\Th(\o)\e \right]} \bigg|_{r=0}^{r = \infty }  = - \frac{2i\Th(\o)}{ (x-y)^2 - i\Th(\o)\e } .
\end{split}
\end{align}
It follows
\begin{align}\label{eq:J-fourier0}
\begin{split}
    \CJ_a(x) &= \frac{1}{2} \bigg( \lim_{\o \to 0^+} + \lim_{\o \to 0^-} \bigg) \p_a \hat q^A(x) \int_\mrr \dt u \int_{\mss^d} \dt^dy \left[ \frac{J_A^+ (u, y ) }{ (x-y)^2 - i\Th(\o)\e }  -  \frac{J_A^- (u, y )}{ (x-y)^2 + i\Th(\o)\e }     \right] \\
    &=  \p_a \hat q^A(x) \int_\mrr \dt u \int_{\mss^d} \dt^dy \, \CP \left( \frac{1}{ (x-y)^2} \right) \big( J_A^+ (u, y )  -   J_A^- (u, y )  \big) , 
\end{split}
\end{align}
where in the last equality, we used the definition of the principal value
\begin{align}
\CP \left( \frac{1}{ (x-y)^2} \right) \equiv \frac{1}{2} \bigg[ \frac{1}{(x-y)^2 + i\e} + \frac{1}{(x-y)^2 - i\e}  \bigg] .
\end{align}
Now, recalling the on-shell momentum parametrization via flat null coordinates given in \eqref{mompar} with $m=0$, we evaluate
\begin{align}\label{eq:integrand0}
\begin{split}
    \p_a \hat q^A(x) J_A^\pm(u,y) &= x_a \big( J_0^\pm(u,y) - J_{d+1}^\pm (u,y) \big) + J_{X^a}^\pm (u,y) ,
\end{split}
\end{align}
where $J^\pm_{X^a}(u,y)$ labels the $a$-component of $J^\pm_A$ in Cartesian coordinates; we use this notation to distinguish it from the $a$-component of $J^\pm_A$ in flat null coordinates, which we denote as usual by $J^\pm_a$. To rewrite the right-hand-side of \eqref{eq:integrand0} in terms of flat null components $J_\mu(u,y)$, we perform the coordinate change
\begin{align}
\begin{split}
J_u(u,r,y) &=  \frac{1}{2} \big( J_{0}(u,r,y) - J_{d+1}(u,r,y) \big) \\
J_a (u,r,y) &= ry_a \big( J_0(u,r,y) - J_{d+1}(u,r,y) \big) + r J_{X^a}(u,r,y) .
\end{split}
\end{align}
It immediately follows that
\begin{align}
\begin{split}
    J_{X^a}(u,r,y) &= \frac{1}{r} J_a(u,r,y) - 2y_a J_u(u,r,y).
\end{split}
\end{align}
Multiplying both sides of the above equations by $|r|^d$ and taking $r \to \pm\infty$, we have
\begin{align}\label{eq:J-coord-transform0}
\begin{split}
    J_u^\pm(u,y) &= \frac{1}{2} \big( J^\pm_{0}(u,y) - J_{d+1}^\pm(u,y) \big),  \qquad J_{X^a}^{\pm}(u,y) = \pm J_a^\pm (u,y) - 2y_a J_u^\pm(u,y), 
\end{split}
\end{align}
where we defined
\begin{align}
	J_a^\pm(u,y) = \lim_{r \to \pm\infty} |r|^{d-1} J_a^\pm(u,r,y) .
\end{align}
Substituting \eqref{eq:J-coord-transform0} into \eqref{eq:integrand0}, we obtain
\begin{align}
    \p_a \hat q^A(x) J_A^\pm (u,y) &= 2(x_a - y_a) J_u^\pm(u,y) \pm J_a^\pm(u,y) .
\end{align}
Substituting this into \eqref{eq:J-fourier0}, we obtain
\begin{align}\label{eq:J-fourier-final0}
\begin{split}
    \CJ_a(x) &= 2 \int_{\mss^d} \dt^d y\, \frac{x_a - y_a }{(x-y)^2}  \int_\mrr \dt u \, \big( J_u^+(u,y) - J_u^-(u,y) \big) \\
    &\qquad + \int_{\mss^d} \dt^d y\,  \frac{1}{(x-y)^2} \int_\mrr \dt u \,\big( J_a^+(u,y) + J_a^-(u,y)  \big) ,
\end{split}
\end{align}
where we have implicitly dropped the principal value notation $\CP$ for simplicity.

We now want to take the shadow transform of \eqref{eq:J-fourier-final0} and then take the divergence. Observing the fact $\CJ_a(x)$ has scaling dimension 1, we compute 
\begin{align}\label{eq:Ja-shadow-div}
\begin{split}
    \frac{1}{2c_{1,1}}\p^a \wt{\CJ_a}(x) &= \frac{1}{c_{1,1}} \p^a_x \int_{\mss^d} \dt ^dz\, \frac{\CI_{ab}(x-z)}{[(x-z)^2]^{d-1}} \int_{\mss^d} \dt^d y  \, \frac{z^b - y^b}{(z-y)^2} \int_\mrr \dt u \, \big( J_u^+(u,y) - J_u^-(u,y) \big)  \\
    &\qquad +  \frac{1}{2c_{1,1}} \p^a_x \int_{\mss^d} \dt^dz \, \frac{\CI_{a}{}^b(x-z)}{[(x-z)^2]^{d-1}} \int_{\mss^d} \dt^d y \, \frac{1}{(z-y)^2} \int_\mrr \dt u \, \big( J_b^+(u,y) + J_b^-(u,y) \big) .
\end{split}
\end{align}
The above equation has to be evaluated with care, since $c_{1,1} = 0$. To do this, we replace $c_{1,1} \to c_{\D,1}$ and $d-1 \to d-\D$, and only take $\D \to 1$ at the end of the calculation. Let us now focus on the second line. Using the identity
\begin{align}\label{eq:id-1}
	\p^a\bigg( \frac{\CI_{ab}(x)}{(x^2)^{d-\D}} \bigg) = \frac{\D-1}{d-\D}\p_b\bigg( \frac{1}{(x^2)^{d-\D}} \bigg),
\end{align}
we compute
\begin{align}\label{eq:term1a}
\begin{split}
    &\frac{1}{2c_{\D,1}} \p^a_x \int_{\mss^d} \dt ^dz \,\frac{\CI_{a}{}^b(x-z)}{[(x-z)^2]^{d-\D}} \int_{\mss^d} \dt^d y \, \frac{1}{(z-y)^2}  \int_\mrr \dt u \, \big( J_b^+(u,y) + J_b^-(u,y) \big)  \\
    &\qquad = \frac{1}{2c_{\D,1}} \frac{\D-1}{d-\D} \p_x^b \int_{\mss^d} \dt^dz\, \frac{1}{[(x-z)^2]^{d-\D}} \int_{\mss^d} \dt^dy \frac{1}{(z-y)^2} \int_{\mrr} \dt u \,\big( J_b^+(u,y) + J_b^-(u,y) \big)   \\
    &\qquad  =  \frac{1}{2c_{\D,1}}\frac{\D-1}{d-\D} \int_{\mss^d} \dt^d z \, \frac{1}{(z^2)^{d-\D}} \int_{\mss^d} \dt^d y \,\frac{1}{(z-y)^2} \int_\mrr \dt u \, \p^b \big( J_b^+(u,y+x) + J_b^-(u,y+x) \big) , 
\end{split}
\end{align}
where in the second equality we set $z \to z+x$ and $y \to y+x$ in the integral and then pulled in the derivative $\p^b_x$. By current conservation, we have $\p^b J_b^\pm(u,x) = 0$.\footnote{This can be proved by taking the large $r$ limit of $\p^A J_A(X) = 0$ and using the fact that $J_r(u,x) = \CO(|r|^{-d-2})$ \cite{He:2019jjk}.} Therefore, only the term involving $J_u^\pm$ on the right-hand-side of \eqref{eq:Ja-shadow-div} survives, and we have 
\begin{align}\label{eq:Ja-shadow-2a}
\begin{split}
    \frac{1}{2c_{1,1}} \p^a \wt{\CJ_a}(x) &= \lim_{\D \to 1} \frac{1}{c_{\D,1}} \p^a_x \int_{\mss^d} \dt^d z \, \frac{\CI_{ab}(x-z)}{[(x-z)^2]^{d-\D}} \int_{\mss^d} \dt^d y \, \frac{z^b - y^b}{(z-y)^2} \int_\mrr \dt u \, \big( J_u^+(u,y) - J_u^-(u,y) \big)  \\
    &= \lim_{\D \to 1} \frac{1}{c_{\D,1}} \int_{\mss^d} \dt^d z \, \frac{\CI_{ab}(z)}{(z^2)^{d-\D}} \int_{\mss^d} \dt^d y  \int_\mrr \dt u\,\p^a_x\p_z^b \ln\big[(z+x-y)^2 \big] \big( J_u^+(u,y) - J_u^-(u,y) \big) .
\end{split}
\end{align}
Using the fact
\begin{align}
    \p^a \p^b \ln ( x^2) = \frac{2\, \CI^{ab}(x)}{x^2},
\end{align}
we can further simplify \eqref{eq:Ja-shadow-2a} to obtain 
\begin{align}
\begin{split}
\int_\mrr \dt u\, \big( J_u^+(u,x) - J_u^-(u,x) \big)  = \frac{1}{2c_{1,1}} \p^a \wt{\CJ_a}(x) ,
\end{split}
\end{align}
which is precisely \eqref{identity}.

\section{Massive Particles}
\label{app:massive}

In Section~\ref{sec:boundary_terms}, we proved that the identity \eqref{eq:codim2-result} holds in the absence of massive particles. To be precise, we ignored the second term in \eqref{FurC_result}. In this appendix, we show that including that term allows us to account for massive particles in the soft effective action. First, we use the fact that in the far future, the only source for the gauge field is the Li\'enard-Wiechert field strength generated by the massive particles (generalized to arbitrary dimensions), so that given a set of massive particles with momenta $p_k^A$ and $U(1)$ charge $Q_k$, the field strength is\footnote{We determined this by covariantizing the four-dimensional Li\'enard-Wiechert field strength (for instance, see Chapter 14 of \cite{Jackson:1998nia}) and then generalizing to all dimensions. The prefactor is fixed by requiring that Maxwell's equations are satisfied for the point-particle current \eqref{JA-explicit} (away from $X^0 = 0$).} 
\begin{equation}
\begin{split}
{\hat F}^C_{AB}(X) &=  \sum_{k\,\text{massive}} \frac{e^2 m_k^d Q_k}{\O_d} \t(\eta_k X^0)  \frac{ p_{kA} X_B - p_{kB} X_A }{ \left[ ( p_k \cdot X )^2 + m_k^2 X^2 \right]^{\frac{1}{2}(d+1)} } ,
\end{split}
\end{equation}
where $\O_d = \frac{2\pi^{(d+1)/2}}{\G((d+1)/2)}$ is the volume of the unit $S^d$, $\t$ the Heaviside function, $\eta_k$ is positive (negative) for outgoing (incoming) particles, and the sum is only over massive particles. The superscript $C$ indicates that this is the Coulombic solution (recall that there is no radiation for the gauge field through $i^\pm$). Using \eqref{mompar} and moving to flat null coordinates, we find
\begin{equation}
\begin{split}
    {\hat F}^C_{ur}(u,r,x) &= \frac{e^2}{2} \sum_{k\,\text{massive}} \frac{2^d m_k^d Q_k}{\O_d} \frac{  \t \big[ \eta_k \big( r ( 1 + x^2 ) + u \big) \big]  \left[ u\o_k - r\o_k \left(   ( x - x_k )^2  + \frac{m_k^2}{\o_k^2}   \right)  \right] }{    \left\{ \o_k^2  \left[ u + r \left( ( x - x_k )^2  + \frac{m_k^2}{\o_k^2} \right)  \right]^2 - 4 u r m_k^2 \right\}^{\frac{1}{2}(d+1)} } .
\end{split}
\end{equation}
Using this, as well as the Legendre duplication formula
\begin{align}
    \G(z) \G\bigg( z + \frac{1}{2}\bigg) = 2^{1-2z} \sqrt{\pi} \G(2z),
\end{align}
the second term in \eqref{FurC_result} evaluates to
\begin{equation}
\begin{split}
 \lim_{u \to \pm\infty} \lim_{r \to \pm\infty} |r|^d {\hat F}^C_{ur}(u,r,x) &=  - \frac{e^2}{2}  \sum_{k \,\text{massive} \atop \eta_k=\pm} Q_k \CK_d ( z_k  , x_k ; x )  , \qquad z_k \equiv \frac{m_k}{|\o_k|} ,
\end{split}
\end{equation}
where $\CK_\D$ is the bulk-to-boundary propagator \eqref{bulk-to-bdy-prop}. Substituting this result into \eqref{term2}, we find that the contribution of massive particles to the boundary action \eqref{Sbdy} is
\begin{align}
    S_{\bdy}[\hat A,\theta] \big|_{\text{massive}} &=  \int_{\mss^d}  \dt^d x \, \t(x)  \sum_{k \,\text{massive}} Q_k \CK_d ( z_k  , x_k ; x ) =  - \frac{1}{2 c_{1,1}} \int_{\mss^d}  \dt^d x \, \wt{C}^a(x) \CJ_a^\text{massive}(x) ,
\end{align}
where $\CJ_a^\text{massive}$ is the soft factor involving only massive particles, and in the last equality we have used the properties \eqref{Kd_property} and \eqref{shadow-identity}. Adding this to \eqref{eq:codim2-result}, we reproduce exactly the massive contribution to \eqref{bdy-claim}.

\bibliography{YMbib}{}
\bibliographystyle{utphys}

\end{document}